\documentclass[aps,preprint,prb,groupedaddress,showpacs]{revtex4}
\usepackage{amsmath,amscd,amssymb,epsfig}

\begin{document}
\bibliographystyle{apsrev}

\title{Rate processes with non-Markovian dynamical disorder}

\author{Igor Goychuk}
\email[]{goychuk@physik.uni-augsburg.de,goychuk@mailaps.org}
\altaffiliation{on leave from Bogolyubov
Institute for Theoretical Physics, Kiev, Ukraine}

\affiliation{Institute of Physics, University of Augsburg,
Universit\"atsstr. 1, D-86135 Augsburg, Germany}

\date{\today}

\begin{abstract}
Rate processes with dynamical disorder are investigated within
a simple framework provided by 
unidirectional electron transfer (ET)
with  fluctuating transfer rate.
The rate fluctuations
are assumed to be described by a {\it non-Markovian} 
stochastic
jump process which reflects conformational dynamics 
of an electron transferring donor-acceptor molecular complex. 
A tractable analytical expression is obtained 
for 
the relaxation  of the donor population (in the 
Laplace-transformed time domain)
averaged over the {\it stationary} conformational fluctuations. The corresponding mean 
transfer time is also obtained in an analytical form. 
The case of two-state fluctuations
is studied in detail for a model incorporating substate diffusion 
within one of the
conformations. It is shown that an increase of the conformational diffusion time
results in a gradual transition from 
the regime of fast modulation characterized by the averaged ET rate to the
regime of quasi-static disorder. This transition occurs at
the conformational mean residence 
time intervals fixed much less than the inverse of the
corresponding ET rates. 
An explanation of this
paradoxical effect is provided.  Moreover, its presence is also
manifested for the simplest, exactly solvable non-Markovian model with
a biexponential distribution of the residence times in one of the conformations.
The nontrivial conditions for this phenomenon to occur are found.   
\end{abstract}

\pacs{05.40.-a,82.20.-w,82.20.Uv}

\maketitle

\section{Introduction}

Rate processes with dynamical disorder are of prominent importance in many areas of
physics and chemistry  \cite{Zwanzig}.
Consider for example a photo-induced electron transfer (ET)
in a donor-acceptor ET complex like bacterial 
photosynthetic center, or various artificial  
donor-bridge-acceptor molecular systems  \cite{JortnerBixon}. 
Light excites
the  electron donating molecule from its ground electronic state, $D$, to the excited state, 
$D^*$. From  $D^*$
the electron is transferred  
to the acceptor molecule $A$, $D^*A\rightarrow D^+A^-$,  
from which it
returns back to the donor in
its ground state. Such ET process can be
repeated many times. If the corresponding energy differences are large,
the backward reactions are negligible. In the Golden Rule approximation, 
the nonadiabatic ET rate is determined primarily by a strong 
electron-vibrational coupling and a weak tunneling coupling 
between the diabatic electronic states. 
It is given by 
$k[\xi]=\frac{2\pi}{\hbar}|V[\xi]|^2 {\rm FC[\xi]}$,  where $V[\xi]$ is
the effective tunneling matrix element and $FC[\xi]$ is the
corresponding 
Franck-Condon factor \cite{JortnerBixon,MayBook}.  
Both $V[\xi]$ and ${\rm FC[\xi]}$ can depend 
in addition on a  conformational
coordinate $\xi(t)$ which reflects stochastic conformational dynamics
of the ET complex. Such dynamics is slow on the time scale
of vibrational relaxation and the rate $k[\xi]$ follows instantaneously to
the conformational changes. Under these assumptions, the ET kinetics
is  described by a simple stochastic relaxation equation 
for the decay of donor population 
\begin{eqnarray}\label{relax}
\dot p_D(t)=-k[\xi(t)] p_D(t)
\end{eqnarray}
with a fluctuating rate $k[\xi(t)]$.
This presents a simplest example of rate processes with  dynamical disorder \cite{Zwanzig}.
 
Similar descriptions have been introduced in ET theory by Agmon and Hopfield
\cite{AgmonHopf} and 
others \cite{others}.  In these studies,
the conformational fluctuations were assumed to
be continuous obeying 
a Smoluchowski equation description. In such a case, the Kramers diffusion
theory approach 
\cite{Kramers,HTB90} provides a standard theoretical framework. 
The conformational fluctuations  
can, however, consist of discrete, abrupt changes. Then,
a discrete state Markovian modeling seems more appropriate 
\cite{HoffmanRatner,GehlMarchChand,GraigeFeherOkamura,GPM95}. 
It should be however remembered that the corresponding discrete conformational states present in reality some
manifolds 
of many substates with a possibly complex internal structure 
\cite{Frauenfelder,MetzKlaftJort99} and the relevant intrinsic dynamics
can produce  non-exponential, broad residence time distributions
(RTDs) of the time intervals $\tau_j$ spent by the molecule 
in the corresponding discrete macrostates $j$. It can be, for example, a  
power law \cite{MetzKlaftJort99}. 
For this reason, an approximate
discrete state description should
generally be {\it non-Markovian}. Similar non-Markovian features, which
are revealed by non-Poissonian statistics 
in the dynamics of single molecules \cite{Wang}, 
came recently in the focus of attention \cite{Xie,Berezh}.

The main goal of this work is to clarify the role of
non-Markovian memory effects for the rate processes with dynamical disorder.
This challenging problem is exemplified by a  two-state 
model incorporating a conformational 
diffusion inside one of the macroconformations. It is shown
that the non-Markovian effects can be of profound importance leading 
to a paradoxical effect. Namely, the averaged ET kinetics can obey a
quasi-static disorder description notwithstanding that the 
ET rate fluctuates
rapidly on the time scale of averaged ET process. 
A resolution of this apparent paradox is provided. Its presence is also
demonstrated in the simplest, exactly solvable non-Markovian model with
a biexponential distribution of the residence time in one of the conformations.
It can be thus of general importance.

\section{Model and theory\label{theory}} 

Let us consider the relaxation dynamics in Eq. (\ref{relax}) with $\xi(t)$ being
a discrete state non-Markovian jump process termed also the
renewal process $\xi(t)$ \cite{Cox}, or continuous time random walk (CTRW) 
\cite{MontrollWeiss,Hughes}.  It is characterized
by the RTDs $\psi_{j}(\tau)$ and the matrix $P_{ij}$ 
of the transition probabilities.  
Namely, the process stays in the state  $j$ for 
some random 
time interval $\tau_j$ (see Fig. \ref{Fig1}) 
distributed in accordance with the probability density $\psi_j(\tau)$
and jumps at the end of each residence time interval into
another state $i$ with the probability $P_{ij}$ ($P_{jj}=0$).
The sum of these probabilities must add to one,
$\sum_i P_{ij}=1$.  Moreover,
the time-intervals are assumed be independently distributed. 
Then, the just given trajectory description (Fig. \ref{Fig1}) 
defines the considered non-Markovian
process completely. Furthermore, one assumes that 
for each given conformation $\xi_j$, 
the corresponding ET rate is constant, $k_j:=k[\xi_j]$.

The task is to average the relaxation dynamics in Eq. (\ref{relax}) over the 
{\it stationary} conformational fluctuations of the ET complex
assuming an initial preparation of the electronic
subsystem always in the excited donor state, i.e., $p_D(0)=1$. 
For the stochastic process $\xi(t)$ {\it prepared} at $t_0=0$ in
a certain nonequilibrium state $j$ a similar problem was solved in Refs.
\cite{Kampen79,ChvostaReinek99}.  However, to arrive at the stationary 
averaging one must perform also an averaging over the stationary initial 
preparations of $\xi(t)$. 
For non-Markovian processes this task is not 
trivial \cite{Cox,Tunaley,Burshtein,Goychuk04}. Indeed,
by preparing the electronic subsystem at $t_0=0$ in one
and the same electronic state with the probability one, $p_D(0)=1$, the whole ET complex will 
be found in a state $j$ with 
the stationary probability $p_j(0)=p_j^{st}=
\lim_{t\to\infty}p_j(t)$. However, this state will be already occupied  
{\it before} $t_0$ for some unknown and random
time $\tau_j^*$. The conditioning on and averaging over this 
time must be done in a stationary setup.
This fact
is well known in the theory of renewal processes \cite{Cox}. 
The averaging over $\tau_j^*$ yields 
a different RTD for the first time sojourn, $\psi^{(1)}_{j}(\tau)=
\int_{\tau}^{\infty}\psi_j(t)dt/\langle \tau_j\rangle$ \cite{Cox},
where 
$\langle \tau_j\rangle = \int_0^{\infty}\tau\psi_j(\tau)d\tau$ 
is the mean residence time (MRT) in the $\xi_j$ conformation.
These MRTs are assumed to be finite in the following.
Only for the strictly exponential 
 RTDs (Markovian case), the RTD for the first sojourn
 $\psi^{(1)}_j(\tau)$ is the same as $\psi_j(\tau)$, i.e., 
 $\psi^{(1)}_j(\tau)=\psi_j(\tau)$.

 A similar problem of {\it stationary averaging} was solved for 
a stochastic Kubo oscillator 
\cite{Kubo62}, 
$\dot x(t)=i\omega[\xi(t)]x(t)$, in Ref. \cite{Goychuk04}.
By observing that the relaxation problem is mathematically equivalent to 
that of the Kubo oscillator
with the imaginary frequency $\omega\to ik$, the use of the results from Ref. 
\cite{Goychuk04} yields for the (Laplace-transformed) stochastically averaged probability
$\langle \tilde p_D(s) \rangle_{st}=\int_0^{\infty}\langle p_D(t) \rangle_{st}\exp(-st)d t$ to occupy the excited donor state the following expression:
\begin{eqnarray}\label{result1}
\langle \tilde p_D(s) \rangle_{st} 
& = 
& \sum_{j}\frac{p_j^{st}}{s+k_j}-
\sum_{j}\frac{1-\tilde \psi_j(s+k_j)}{(s+k_j)^2}
\frac{p_j^{st}}{\langle \tau_j\rangle } \nonumber \\ 
& + &
\sum_{n,l,m}\frac{1-\tilde \psi_l(s+k_l)}{s+k_l}
\left (\frac{1}{I-P\tilde D(s)}\right )_{lm}
\\
&\times &P_{mn}
\frac{1-\tilde \psi_n(s+k_n)}
{s+k_n}\frac{p_n^{st}}{\langle \tau_n\rangle }\;, \nonumber
\end{eqnarray}
where the stationary occupational probabilities $p_j^{st}$ of the conformational states $\xi_j$ are the solutions of
\begin{eqnarray}
\frac{p_i^{st}}{\langle \tau_i\rangle }=\sum_j P_{ij}
\frac{p_j^{st}}{\langle \tau_j\rangle }.
\end{eqnarray}
In Eq. (\ref{result1}), $\tilde D_{nm}(s)=\delta_{nm}\tilde \psi_m(s+k_m)$, $I$ is the unity matrix
and
$\tilde f(s):=\int_0^{\infty}e^{-st}f(t)dt$ denotes the
Laplace-transform \cite{Korn}  of any corresponding function $f(t)$.

The result in Eq. (\ref{result1}) provides the formally {\it exact} 
solution of the 
studied problem. The first sum in Eq. (\ref{result1}) presents
the solution of the problem with  frozen static disorder. It can be
trivially inverted to the time-domain:
\begin{eqnarray}\label{static}
\langle p_D(t)\rangle_{static} = \sum_{j=1}^{N}
p_j^{st}\exp(- k_j t)\;.
\end{eqnarray}
The remaining part presents nontrivial dynamical contributions.
Its analytical inversion to the time domain is possible in a few
exceptional cases only. However, in many cases it can reliable 
be inverted numerically by use of 
the Stehfest method  \cite{Stehfest}.  Moreover,
the exact analytical expression for the mean survival (transfer) time
\cite{AgmonHopf,others}, 
$\tau_{ET}:=\int_{0}^{\infty}\langle p_D(t)
\rangle dt$, follows  as $\tau_{ET}=
\langle \tilde p_D(0)\rangle_{st}$. It can be readily given
in an explicit form for ergodic processes with uniform mixing, 
where $p_j^{st}=\langle \tau_j\rangle/\sum_j\langle 
\tau_j\rangle$ and $\sum_j P_{ij}=1$ \cite{Goychuk04}.  
This is also the case of two-state conformational fluctuations
with $P_{12}=P_{21}=1$ (alternating renewal process \cite{Cox}) which is investigated below.

\section{ Two-state conformational fluctuations}

In this simplest case which is of major interest in applications, the
general result in Eq. (\ref{result1}) reduces to:
\begin{eqnarray}\label{result2}
& &\langle \tilde p_D(s)\rangle_{st}  =   
\frac{1}{\langle \tau_1 \rangle+
\langle \tau_2 \rangle}\Big \{ \sum_{j=1,2}
\frac{\langle \tau_j \rangle}{s+k_j}
   \\ 
& - & \frac{(k_1-k_2)^2}{(s+k_1)^2(s+k_2)^2} 
\frac{[1-\tilde \psi_1(s+k_1)][1-\tilde \psi_2(s+k_2)]}
{1-\tilde \psi_1(s+k_1)\tilde \psi_2(s+k_2)}\Big \}\;. \nonumber 
\end{eqnarray}
This result can also be obtained from one in Ref. \cite{Jung} by virtue of
the above formal analogy with the Kubo oscillator. The corresponding exact expression
for the mean transfer time  reads:
\begin{eqnarray}\label{result3}
\tau_{ET}  & = &  \frac{1}{\langle \tau_1 \rangle+
\langle \tau_2 \rangle}  \Big \{ \frac{\langle \tau_1 \rangle}{k_1}
+\frac{\langle \tau_2 \rangle}{k_2}\nonumber \\
  & - &  \left(\frac{1}{k_2}- \frac{1}{k_1}\right)^2
\frac{[1-\tilde \psi_1(k_1)][1-\tilde \psi_2(k_2)]}
{1-\tilde \psi_1(k_1)\tilde \psi_2(k_2)}\Big \}\;. 
\end{eqnarray}
This expression is valid for arbitrary RTDs with finite MRTs and presents one
of important results. It defines also an effective ET rate as
$\Gamma:=\tau_{ET}^{-1}$.

\subsection{ Markovian fluctuations}

Let us first reproduce the known results for the case of Markovian fluctuations from the above 
general formal
expressions. For a Markovian process, the both RTDs are strictly  
exponential: $\psi_{1,2}(\tau)=\langle \tau_{1,2}\rangle^{-1}\exp(-\tau/\langle \tau_{1,2}\rangle)$.
The trajectory description with such RTDs
is equivalent to a rate description 
with  the conformational transition rates $\gamma_{1,2}=
\langle \tau_{1,2}\rangle^{-1}$. The corresponding Laplace-transformed RTDs  read $\tilde \psi_{1,2}(s)=1/(1+s\langle \tau_{1,2}\rangle)$.
In this simplest case, the inversion 
of Eq. (\ref{result2}) to the time domain
yields a biexponential relaxation,
\begin{eqnarray}\label{Mark}
\langle p_D(t)\rangle_{st}  = 
c_1 \exp(-\Gamma_1t)+ c_2 \exp(-\Gamma_2t)\,,
\end{eqnarray}
with the relaxation (transfer) rates
\begin{eqnarray}\label{relrate}
\Gamma_{1,2}=\frac{1}{2}\{ k_1+k_2 +\gamma_1 +\gamma_2 
\mp\sqrt{(k_1+\gamma_1-k_2 -\gamma_2)^2+4\gamma_1\gamma_2} \}
\end{eqnarray} 
 and the weight coefficients (probabilities of different relaxation channels)
\begin{eqnarray} \label{coef}
c_{1,2}=\frac{1}{2}\left\{ 1 \pm
\frac{\gamma_1+\gamma_2+(k_1-k_2)(\gamma_1-\gamma_2)/(\gamma_1+\gamma_2)}
{\sqrt{(k_1+\gamma_1-k_2 -\gamma_2)^2+4\gamma_1\gamma_2}}\right\}. 
\end{eqnarray}
The corresponding mean transfer time reads
\begin{eqnarray}\label{MET}
\tau_{ET}=\frac{k_1\gamma_1+k_2\gamma_2+(\gamma_1+\gamma_2)^2}{(\gamma_1+\gamma_2)
(k_1k_2+k_1\gamma_2+k_2\gamma_1)}\,.
\end{eqnarray}
The result in Eqs. (\ref{Mark})-(\ref{coef})  reproduces, e.g.,
the known result in Ref. \cite{GehlMarchChand} for the case
of identical transition rates $\gamma_1=\gamma_2=\gamma$. This latter one
can also be obtained from the well-known
solution of the Kubo oscillator \cite{Kubo62,KampenBook} by virtue of
the above mentioned formal analogy.
The remarkably simple result in Eqs. (\ref{Mark})-(\ref{coef}) captures in essence three 
different transport regimes with dynamical disorder. If the conformational dynamics is slow,
$\gamma_1\gamma_2\ll (k_1+\gamma_1-k_2 -\gamma_2)^2/4$, then the limit 
in Eq. (\ref{static}) is approximately 
reproduced with $p_{1,2}^{st}=\langle \tau_{1,2}\rangle/(\langle \tau_{1}\rangle+\langle \tau_{2}\rangle)$ (the regime of quasi-static disorder). If $\gamma_{1,2}\gg 
k_{1,2}$ (the fast modulation regime), then ET is well described by the averaged rate $\langle 
k\rangle=p_1^{st}k_1+p_2^{st}k_2$, i.e.,
\begin{eqnarray}
\langle p_D(t)\rangle_{st} \approx \exp(-\langle k\rangle t).
\end{eqnarray}
One more interesting regime occurs for
$\gamma_1=\gamma_2=\gamma$ when 
$\frac{k_1 k_2}{k_1+k_2}\ll \gamma\ll \langle k\rangle/2$.
In this case, the mean 
transfer time $\tau_{ET}$ equals approximately to the autocorrelation time
$\tau_{corr}=1/(2\gamma)$ of the Markovian two-state fluctuations, as it can be deduced
from Eq. (\ref{MET}), i.e.
\begin{eqnarray}\label{gated}
\tau_{ET}\approx  \tau_{corr}.
\end{eqnarray}
In other words, ET becomes {\it gated},
i.e., it is chiefly  controlled by conformational fluctuations 
\cite{McCammon,HoffmanRatner,GraigeFeherOkamura}. 
An interesting gated regime emerges in the case where both ET rates are very different,
$k_2\ll k_1$, and the ET complex stays most of the time in the weakly conducting conformation, i.e.,
$\langle \tau_{2}\rangle\gg \langle \tau_{1}\rangle$. Then, for $k_2\ll \gamma_{2}\ll\gamma_{1} \ll k_1$ 
Eq. (\ref{MET})
yields 
\begin{eqnarray}\label{gated2}
\tau_{ET}\approx \langle \tau_2\rangle\,.
\end{eqnarray}
At the first look, the result in Eq. (\ref{gated2}) might seem puzzling: the mean transfer time
is given by the mean time intervals spent in the {\it weakly} conducting conformation. It is, however, quite natural
since $\gamma_2=\langle \tau_2\rangle^{-1}$ is nothing but the switching-on rate which corresponds to
the rate limiting step of the overall ET process.

\subsection{\label{diffusionmodel}Non-Markovian fluctuations}

To demonstrate a new surprising effect
due to a broad non-exponential distribution of the residence times, 
let us 
consider the following tractable model. The conformation with 
the larger ET rate $k_1$
(the ``on''-conformation) has an 
exponential distribution of
residence times with MRT $\langle \tau_1\rangle$. In contrast, 
the ``off''-conformation with a much smaller rate $k_2\ll k_1$ is supposed 
to consist of a large number of  quasi-degenerate subconformations. Correspondingly, it can undergo
a (sub)diffusion among these subconformations, which all possess one and the same ET rate $k_2$, with a 
characteristic diffusional time $\tau_d$.
This time can be related to the volume of the corresponding conformational subspace (modeled
here as an one-dimensional box, cf. Appendix \ref{Appendix1}), or to the number of quasi-degenerate diffusional substates, i.e. to the statistical weight of the conformation ``2''. For this reason,  
within a statistical thermodynamic interpretation this
time can also be related to the entropy of the conformational macrostate ``2''. 
Furthermore, one assumes that a conformational rate transition can occur into the ``on''-conformation
by crossing one of the boundaries of the underlying quasi-one-dimensional
conformational subspace.
An effective rate of such transition is $\gamma_2:=\langle \tau_2\rangle^{-1}$. 
The corresponding
diffusional model, which is outlined in more detail in the Appendix \ref{Appendix1},  
is characterized by the Laplace-transformed RTD \cite{PNAS02,new}
\begin{eqnarray}\label{distribution}
\tilde \psi_2(s) = \frac{1}{1 + s\langle \tau_2\rangle 
g_{\alpha}(s\tau_d) }\;,
\end{eqnarray}  
where 
\begin{eqnarray}
g_{\alpha}(z)=\frac{ \tanh(z^{\alpha/2})}{z^{\alpha/2}}.
\end{eqnarray}
The function $g_{\alpha}(z)$ one has the following remarkable properties: $g_{\alpha}(0)=1$ and $g_{\alpha}(z)$ 
decays monotonously to zero with increasing $z$. The case 
$\alpha=1$ corresponds to the normal diffusion \cite{PNAS02} 
and $0<\alpha < 1$ corresponds to
a subdiffusion with index $\alpha$ \cite{new}. The full details of the derivation
of Eq. (\ref{distribution}) and the relevant discussion
are given in Ref. \cite{new}. 

Let us consider first the case of normal diffusion, $\alpha=1$.
For $\tau_d=0$, Eq. (\ref{distribution}) yields a single-exponential RTD. 
For $\tau_d \ll 
\langle \tau_2\rangle$, the intra-conformational diffusion effects are
not essential. However, for $\tau_d \gg \langle \tau_2\rangle $
the situation changes. Namely, for  
$\langle \tau_2\rangle^2/\tau_d\ll\tau\ll \tau_d$
the corresponding RTD becomes a negative power law, 
$\psi_2(\tau)\propto \tau^{-3/2}$, which
ends by an exponential tail for $\tau>\tau_d$ \cite{PNAS02}.
The used model allows one to trace  
the influence of non-Markovian
effects on the averaged transport process by assuming different variable values of 
$\tau_d$ with other model parameters, $k_1,k_2,\langle \tau_1\rangle,\langle \tau_2\rangle$, 
being fixed and defining the corresponding Markovian reference model. For this goal, we consider
below a transfer regime with $k_1\langle \tau_1\rangle\ll 1$, 
$k_2\langle \tau_2\rangle\ll 1$ which must be regarded as a 
fast modulation regime 
in the Markovian case. For $k_1\langle \tau_1\rangle\ll 1$, Eq. (\ref{result3}) yields
the following approximate 
analytical expression for $\tau_{ET}$: 
\begin{eqnarray}\label{result40}
\tau_{ET}\approx\frac{1}{k_1k_2(\langle \tau_1\rangle+\langle \tau_2\rangle)}\left( 
k_1\langle \tau_2\rangle+k_2\langle \tau_1\rangle-\frac{(k_1-k_2)^2
\langle \tau_1\rangle\langle \tau_2\rangle
g_{\alpha}(k_2\tau_d)}{k_1\langle \tau_1\rangle+k_2\langle \tau_2\rangle g_{\alpha}(k_2\tau_d)}
 \right). 
\end{eqnarray}
Interestingly enough, for 
$\langle \tau_1\rangle=\langle \tau_2\rangle=\langle \tau\rangle$, 
the dependence on $\langle \tau\rangle$ in Eq. (\ref{result40})
drops out and it becomes
\begin{eqnarray}\label{result4}
\tau_{ET}\approx\frac{1}{2k_1k_2}\left( 
k_1+k_2-\frac{(k_1-k_2)^2g_{\alpha}(k_2\tau_d)}{k_1+k_2g_{\alpha}(k_2\tau_d)}
 \right).
 \end{eqnarray}
The result in Eq. (\ref{result40}) allows one to trace {\it analytically} 
the emergence of the regime of 
quasi-static disorder with the increase of $\tau_d$. Namely, for
$\tau_d=0$, $g_{\alpha}(0)=1$ and Eq. (\ref{result40}) yields
$\tau_{ET}\approx \langle k\rangle^{-1} $ which corresponds to the averaged 
rate description in accordance with the intuitive expectations.  However,
if
$\tau_d\to\infty$, then $g_{\alpha}(k_2\tau_d)\to 0$ and 
$\tau_{ET}=p_1^{st}k_1^{-1}+
p_2^{st}k_2^{-1}$. This corresponds to the quasi-static regime.  
Therefore, a transition to the quasi-static regime can occur due to an increase of the
conformational diffusion time $\tau_d$ even if the MRTs $\langle \tau_1\rangle$ and
$\langle \tau_2\rangle$ are kept fixed
well within the Markovian fast modulation regime. The corresponding criterion reads:
\begin{eqnarray}\label{criterion}
 g_{\alpha}(k_2\tau_d)\ll 1.
\end{eqnarray}
This surprising result is corroborated by the numerical
inversion of Eq. (\ref{result2}) depicted  
in Fig. \ref{Fig2}
for the following set of parameters:
$k_1=1,\; k_2=0.1,\;\langle \tau_1\rangle=
\langle \tau_2\rangle=0.1 $ with varying $\tau_d$.
The numerical inversion of the Laplace-transform (\ref{result2}) was done using
the improved Stehfest method \cite{Stehfest,Valko} following a ``rule of thumb''
found empirically: the number $N$ of terms 
used in the corresponding asymptotic series expansion \cite{Stehfest} ($N=16$ in the original and
standard version)
should correspond to the
digital precision  of calculations \cite{Valko}. 
The latter one (along with the number of terms in the Stehfest series) 
was made variable (increased)
until the convergency of the numerical results was reached. This was made
possible
by implementing
the Stehfest algorithm using the Computer Algebra System MAPLE which allows
for numerical calculations with an arbitrary precision. $N=32$
was sufficient for the numerical calculations done in this paper to arrive
at the required accuracy (within the line width in the figures representing the numerical
results).  
 Note that the mean residence times $\langle 
\tau_{1,2}\rangle$, 
which are used for calculations in Fig. \ref{Fig2}, 
are indeed much less than the inverse of the corresponding ET rates. 
This relates
also to the mean turnover time, 
$\tau_{turn}:=\langle 
\tau_{1}\rangle+\langle \tau_{2}\rangle\ll k_1^{-1},k_2^{-1}$.
Therefore, one intuitively expects that many stochastic turnovers
should occur before the electron donor population changes significantly.
This in turn should validate an average rate description of the ET process.  
Contrary to this intuitively clear picture which is well established
in the case of Markovian fluctuations \cite{Zwanzig}, an increase of $\tau_d$
results in  a gradual transition to the regime of quasi-static disorder,
see in Fig. \ref{Fig2}. This presents certainly a paradox: Judging
from the time scale $\tau_{turn}$ the ET complex fluctuates
rapidly (on the time scale of decay of the donor population) 
between two conformations.
Nevertheless,  the averaged rate description breaks down and 
for large $\tau_d$ the averaged ET process becomes 
non-exponential entering a slow modulation regime. 
It can be even better described by the quasi-static 
approximation (top curve in Fig. \ref{Fig2}) which corresponds to
a ``frozen" disorder. 

A deceivingly clear 
resolution of this paradox is based on 
the observation 
that for the considered non-Markovian process the effective autocorrelation
time can be much larger than the autocorrelation time of its Markovian
counterpart with the same MRTs \cite{GoychukHanggi03/04}.
The mean autocorrelation time $\tau_{corr}$ can serve as a quantifier for memory effects.
It can be defined as
$\tau_{corr}:=\int_0^{\infty}|cov(t)|dt$, where $cov(t):=\langle \delta k(t)\delta k(0)\rangle_{st}/
\langle \delta k^2 \rangle_{st}$ is the normalized
autocorrelation function of $k(t)$ (here, $\delta k(t):=k(t)-\langle k\rangle$) 
and obtained by using the Stratonovich formula \cite{Stratonovich,Wang,
GoychukHanggi03/04}
\begin{eqnarray}\label{laplace-corr}
\overline{cov}(s)=\frac{1}{s}-\left(\frac{1}{\langle \tau_1\rangle} +
\frac{1}{\langle \tau_2\rangle}\right)\frac{1}{s^2}\frac{\left(1-\tilde \psi_1(s)\right)
\left(1-\tilde \psi_2(s)\right)}{\left(1-\tilde \psi_1(s)\tilde
\psi_2(s)\right)}
\end{eqnarray}
for the Laplace-transform of the normalized autocorrelation function of a stationary two-state
renewal process. If $cov(t)$ does not change its sign in time (the present situation),
then $\tau_{corr}=\lim_{s\to  0}\overline{cov}(s)$ and by use of
Eq. \ref{laplace-corr} the mean autocorrelation time 
 reads \cite{GoychukHanggi03/04}:
\begin{eqnarray}\label{corr}
\tau_{corr}=\frac{1}{2}(C_1^2+C_2^2)\tau_M
\end{eqnarray}
where 
\begin{eqnarray}\label{cv}
C_j=\frac{\sqrt{
\langle \tau_j^2\rangle - \langle \tau_j\rangle^2}}{
\langle \tau_j\rangle}
\end{eqnarray}
 is the coefficient of variation of 
the corresponding RTD $\psi_j(\tau)$ and 
\begin{eqnarray}\label{tauM}
\tau_M=\frac{\langle \tau_1\rangle
\langle \tau_2\rangle}{\langle \tau_1\rangle +
\langle \tau_2\rangle}
\end{eqnarray}
 is the autocorrelation time of the Markovian
process with the same MRTs $\langle \tau_1\rangle$ and $\langle \tau_2\rangle$.
For the considered model, $
\tau_{corr}=
(1+ \tau_d/3\langle \tau_2\rangle)\tau_M$. 
It can be much larger than $\tau_M$ 
with the increase
of $\tau_d$. In contrast with 
the Markovian process,
the time-correlations do not decay on the time scale of stochastic
turnovers
and can persist very long. This can be considered as a reason for the 
transition to the quasi-static regime.
Judging from
the autocorrelation time scale it can indeed be the quasi-static regime. 

Accepting this reasoning uncritically,  one might conclude that
for the processes with a {\it divergent} RTD variance, $C_{2}\to \infty$,
and thus with 
$\tau_{corr}\to\infty$, 
the quasi-static regime will be realized always. However, such a conclusion
is not correct. 
Indeed, let us take some $\alpha<1$ in 
Eq. (\ref{distribution}) (this corresponds to a subdiffusion with index $\alpha$ \cite{new}). 
Then, it is easy to show that formally $\tau_{corr}=\infty$ for all $\tau_d\neq 0$ 
and one might naively expect that
the regime of quasi-static disorder in the ET kinetics will readily be realized.
However, the numerics  do not support this expectation.
For example, for $\alpha=0.1$ and for $\tau_d=100$ 
the averaged ET is not exponential indeed. Nevertheless, it is closer to the
regime of fast modulation rather than to the regime of
quasi-static disorder. This non-occurrence of the regime of
quasi-static disorder
agrees, however, with the criterion (\ref{criterion})
which yields in this case $g_{\alpha}(k_2\tau_d)\approx 0.72$ 
for the given parameters. This value is indeed closer to
$g_{\alpha}(k_2\tau_d)=1$ (the averaged rate description) rather than
to $g_{\alpha}(k_2\tau_d)=0$ (the regime of quasi-static disorder).

It is {\it not} the
divergence of the mean autocorrelation time which does play a crucial
role, 
but an intermittent, bursting
character of non-Markovian fluctuations which is 
important for the discussed phenomenon to occur. 
For $\tau_d=100$ in Fig. \ref{Fig2}, the conformational 
fluctuations occur by strongly correlated bursts separated by 
long interburst periods as revealed by
the stochastic simulations of $k(t)$ (see in Fig. \ref{Fig5}(a),(b)
where this striking feature is presented for a different model).
Namely, even if the mean residence time $\langle \tau_2\rangle$ 
in the weakly 
conducting conformation  is very short, $\langle \tau_2\rangle\ll k_2^{-1}$,
the ET complex can nevertheless be trapped occasionally in this conformation
for a very long time period of the order of $\tau_d\gg k_2^{-1}$.
The time $\tau_d$ is a characteristic time which is required to
explore by random walk the manifold of quasi-degenerate 
conformational substates 
within the corresponding macroconformation. Such long time intervals separate the
bursts within which the ET complex fluctuates rapidly between its two 
macroconformations. The mean residence time $\langle \tau_2'\rangle$ of the 
time intervals spent in the conformation $\xi_2$ (conformation ``2'') within a burst
could obviously be much less than the overall $\langle \tau_2\rangle$ (since
this latter one
accounts also for the long interburst sojourn in the same conformation),
i.e. $\langle \tau_2'\rangle\ll \langle \tau_2\rangle$. For
 $\langle \tau_1\rangle =\langle \tau_2\rangle$ and $p_1^{st}=p_2^{st}$, this means that during the
the burst duration the ET complex spends most of the time in the conformation ``1''
since $\langle \tau_1\rangle\gg \langle \tau_2'\rangle$. 
If the burst durations $\langle \tau_{burst}\rangle$ are, 
on average, much larger than the 
inverse ET rate $k_1^{-1}$,  i.e. $\langle \tau_{burst}\rangle\gg k_1^{-1}$,
 then it becomes clear why the quasi-static
averaging does apply in such a situation, even if 
$\langle \tau_1\rangle =\langle \tau_2\rangle\ll k_1^{-1}, k_2^{-1}$.
This is the reason why the averaging over 
many repeated realizations of electron transfer in 
the same ET complex, or
an ensemble averaging for a number of such complexes done at the same time can indeed 
yield a quasi-static regime even if a large portion of these complexes does fluctuate
very fast on the time scale of ET while another one stays temporally ``frozen'' in
the ``off''-conformation. 
However, even if $\tau_{corr}$
is formally infinite ($\alpha<1$), the fluctuations do not
necessarily have a distinct bursting character. Then, the discussed
phenomenon does not necessarily occur.  This qualitative explanation
is fully confirmed within the following {\it exactly} solvable non-Markovian model.

\subsection{\label{exactmodel} Exactly solvable non-Markovian model}

The simplest non-Markovian model is obtained by assuming a biexponential RTD,
\begin{eqnarray}\label{biexp}
\psi_2(\tau)=\theta\alpha_1\exp(-\alpha_1 \tau)+ (1-\theta)\alpha_2\exp(-\alpha_2 \tau) ,
\end{eqnarray}
where $\alpha_{1,2}$ are two different transition rates for the conformational transition
from the conformation ``2'' to
the conformation ``1''. They can be realized with the probabilities 
$0<\theta<1$ and $1-\theta$, correspondingly. The corresponding mean residence time is
 \begin{eqnarray}\label{MRT2}
\langle \tau_2\rangle=\theta /\alpha_1+(1-\theta )/\alpha_2,
\end{eqnarray}
and the coefficient of variation (\ref{cv}) for this distribution is
\begin{eqnarray}
C_2(\theta,\zeta)=\sqrt{2\frac{f_1(\theta,\zeta)}{f_2(\theta,\zeta)}-1}, 
\end{eqnarray}
where the functions $f_{1,2}(\theta,\zeta)$ are given in Eq. (\ref{aux2}) and $\zeta=\alpha_1/\alpha_2$
is the ratio of two rate constants in Eq. (\ref{biexp}). For $\zeta\ll 1$ and $\zeta \ll \theta$ the following approximate expression is valid
\begin{eqnarray}\label{C2approx}
C_2(\theta,\zeta)\approx \sqrt{\frac{2}{\theta}-1}\;.
\end{eqnarray}
Note that $C_2$ can be very large. This applies when
$\theta\ll 1$ and $\zeta\ll 1$,  see in Fig. \ref{Fig3}.
This is the case
of distributions (\ref{biexp}) possessing a very small, but very broad tail. Furthermore,
the Laplace-transform in Eq. (\ref{laplace-corr})
can be inverted analytically for this model to obtain a biexponentially decaying autocorrelation
function $cov(t)$. In the parameter regime corresponding to Eq. (\ref{C2approx}), this autocorrelation
function becomes, however, approximately single
exponential,
\begin{eqnarray}\label{cov2}
cov(t)\approx \exp(-\theta t/\tau_M),
\end{eqnarray}
 with $\tau_M$ in Eq. (\ref{tauM}). The result in Eq. (\ref{cov2}) becomes exact in the formal limit
 $\alpha_2\to \infty$, where $\psi_2(\tau)\to \theta\alpha_1\exp(-\alpha_1\tau)+(1-\theta)\delta(\tau)$ and 
 $\langle \tau_2\rangle \to \theta/\alpha_1$. This striking feature must be emphasized: the exists a whole
 family of non-Markovian two-state processes with an {\it exponentially} decaying autocorrelation
 function. The autocorrelation time of these processes is, however, enlarged by the factor $1/\theta$
 as compare with their Markovian counterpart possessing the same MRTs $\langle \tau_1\rangle$
 and $\langle \tau_2\rangle$.
In accordance with Eq. (\ref{cov2}), the autocorrelation time of such non-Markovian processes
$\tau_M/\theta$
can be much larger than the autocorrelation time $\tau_M$ of their Markovian counterpart.

The considered two-state non-Markovian process can also be embedded into a three state
Markovian kinetic scheme with two substates (subconformations) corresponding to the conformation ``2''.
The two-state fluctuations of the ET rate $k(t)$ present nevertheless a non-Markovian process.
The just described model can be solved exactly. 
Namely, Eq. (\ref{result2}) yields after some lengthy calculations a rational function
\begin{eqnarray}\label{res1biexp}
\langle \tilde p_D(s)\rangle_{st}=\frac{s^2+a_1s+a_0}{s^3+b_2s^2+b_1 s +b_0},
\end{eqnarray}
where the polynomial coefficients $a_j,b_i$ are given in the Appendix \ref{Appendix2}.
The analytical inversion of  Eq. (\ref{res1biexp}) to
the time domain yields formally a three exponential kinetics
 \begin{eqnarray}\label{3evolv}
\langle p_D(t)\rangle_{st} =\sum_{i=1,2,3} c_i \exp(-\Gamma_i t)
\end{eqnarray}
with the rates $\Gamma_i$ which are given in the Appendix \ref{Appendix2}, Eqs. 
(\ref{3rates})-(\ref{lastrate}), and the weights $c_i=(\Gamma_i^2-a_1\Gamma_i+a_0)/(3\Gamma_i^2-2b_2\Gamma_i+b_1)$.
A numerical evaluation for all cases studied in this paper shows, however, that the ET kinetics remains
 effectively
 biexponential (at most) since one of the rates $\Gamma_2$ exceeds much the other two and the
corresponding weight $c_2$ is negligible small.

In order to establish a criterion similar to Eq. (\ref{criterion}), the Laplace-transform of (\ref{biexp}) 
can be represented in a form similar to Eq.
(\ref{distribution}),
\begin{eqnarray}\label{distribution2}
\tilde \psi_2(s) = \frac{1}{1 + s\langle \tau_2\rangle 
g\left (s\langle \tau_2\rangle,\theta,\zeta\right) }\;,
\end{eqnarray}  
where  
\begin{eqnarray}\label{gnew}
g\left(z,\theta,\zeta\right)= \frac{[\theta+(1-\theta)\zeta]^2+\zeta z}{[\theta+(1-\theta)\zeta]^2+
[\theta+(1-\theta)\zeta][1-\theta+\theta \zeta]z},
\end{eqnarray}
and MRT $\langle \tau_2\rangle$ is given in Eq. (\ref{MRT2}). Then, the criterion (\ref{criterion})
is get modified for the present model as
\begin{eqnarray}\label{criterion2}
 g\left(k_2\langle \tau_2\rangle,\theta,\zeta\right)\ll 1
\end{eqnarray}
with $k_2\langle \tau_2\rangle\ll 1$. Let us clarify if it is possible to fulfill
the criterion (\ref{criterion2}). It is rather nontrivial
in the present case since we do not have such an independent parameter as the diffusional time
$\tau_d$   which can easily be adjusted to satisfy the condition (\ref{criterion}) for a sufficiently
large value of $\tau_d$. Therefore, a deliberate choice of parameters is required. 
 We consider the following set of parameters $k_1=1, k_2=0.2$, and 
 $\langle \tau_1\rangle=\langle \tau_2\rangle=0.2$ to be still within a fast modulation
 regime in accordance with the Markovian criteria. The plot of the function $g\left(z,\theta,\zeta\right)$
versus $\theta$ at $z=0.04$ and at the different values of $\zeta$ in Fig. \ref{Fig4} reveals
that the criterion (\ref{criterion2}) can indeed be fullfield for a very small ratio of
the rate constants, starting from
$\zeta=10^{-3}$, and for a very small, but finite splitting probability $\theta$ 
(note the presence of an
optimal value of $\theta$ in Fig. \ref{Fig4}, where $g\left(k_2\langle \tau_2\rangle,\theta,\zeta\right)$
reaches a conditional minimum).  
The corresponding conformational dynamics is
profoundly bursting indeed, with long time intervals between bursts
as it is revealed by the stochastic simulations shown in the Fig. \ref{Fig5} (a),(b). Here, the interburst intervals have the characteristic
time scale $\tau_{inter}\approx \langle\tau_2\rangle /\theta=20>
k_{1,2}^{-1}$. The burst durations have
about the same characteristic time scale and within the burst the excursions into
the state ``2'' are extremally brief and the process stays mostly in the state ``1''. 
These are the reasons why a regime similar to the regime of quasi-static disorder is realized.
In the case $\theta=0.5$ (which is close to the regime of fast modulation, cf. Fig. \ref{Fig6}), 
$k(t)$ has no such profoundly bursting character -- compare the Fig. 5(c),
where a sample trajectory of this non-Markovian process is depicted, with the Fig. 5(a),(b) and 
with the Fig. 5(d),
where a sample trajectory of the corresponding Markovian process is shown. Clearly, in this case
$k(t)$ reminds strongly its Markovian counterpart. This similarity can be, however, somewhat deceiving.
The sample trajectories in Fig. 5(c) and Fig. 5(d) are replotted in Fig. 5(e) and Fig. 5(f), respectively,
on a much shorter time scale. The both processes have the same $\langle \tau_{1}\rangle=\langle \tau_{2}\rangle=0.2$. However, the trajectory in Fig. 5(e) corresponds to a non-Markovian process, where
some very brief excursions occurs into the state ``2'' during the sojourn in the state ``1''. This non-Markovian process has, in accordance with Eq. (\ref{cov2}),
the autocorrelation time which is twice bigger as compare with its Markovian counterpart in Fig. 5(f).

The stationary stochastic trajectories in Fig. \ref{Fig5} were generated due to the following algorithm.
First, a random number $0<y_1<1$ is drawn from the uniform distribution \cite{Press} and one decides in
which conformation to start. Namely, if $y_1\leq p_1^{st}$, then one starts in the
state ``1'', otherwise
the state ``2'' is taken as initial. 
If the initial value is $k_1$,  then a random residence time-interval $\tau_1$ 
is drawn from the corresponding
exponential distribution. Here, in accordance with the rule of transformation of random
variables,  $\tau_1=\langle \tau_1\rangle\ln(1/y)$, where $y$ is a new random number taken from 
the uniform
distribution \cite{KampenBook,Press}. After this, the process $k(t)$ flips into the state ``2''
and two new random numbers $y_1$ and $y_2$ are drawn from the uniform 
distribution. $y_1$ is used to decide in accordance
with which exponential distribution $\psi_{21}(\tau)=\alpha_1\exp(-\alpha_1\tau)$, or 
$\psi_{22}(\tau)=\alpha_2\exp(-\alpha_2\tau)$ the process will sojourn in the state ``2''. If $y_1\leq \theta$, then it resides in the state ``2'' in accordance with the distribution $\psi_{21}(\tau)$, i.e.,
$\tau_2=\ln(1/y_2)/\alpha_1$, otherwise 
the distribution $\psi_{22}(\tau)$ is used, i.e.,
$\tau_2=\ln(1/y_2)/\alpha_2$. After the sojourn of duration $\tau_2$ in the state ``2'', the
process flips into the state ``1'' and a new random number $y$ is generated to obtain the
residence interval $\tau_1=\langle \tau_1\rangle\ln(1/y)$. The whole procedure repeats. If the
initial state happens to be the state ``2'' (cf. Fig. 5(b)), then the whole procedure is modified.
Namely, for the {\it first } sojourn in the state ``2'', a different 
RTD $\psi_2^{(1)}(\tau)=\int_{\tau}^{\infty}\psi_2(t)dt/\langle \tau_2\rangle$ is used 
(see the discussion in Sec. \ref{theory}) which is  
\begin{eqnarray}\label{biexp2}
\psi_2(\tau)=\theta_1\alpha_1\exp(-\alpha_1 \tau)+ (1-\theta_1)\alpha_2\exp(-\alpha_2 \tau),
\end{eqnarray}
with 
\begin{eqnarray}\label{theta1}
\theta_1=\frac{\theta}{\theta+(1-\theta)\zeta}.
\end{eqnarray}
This modification is {\it very important}. The generated stochastic trajectories 
will not be {\it stationary} otherwise. As a matter of  fact, $\theta_1$ can be even closer to
one when $\theta\ll 1$, but $\zeta\ll \theta $. The stationary probability to find the considered ET
complex initially in the ``off``-subconformation with the transition rate 
$\alpha_1\ll \alpha_2$
is $p_2^{st}\theta_1$, and not  $p_2^{st}\theta$! $\theta$ is in fact the splitting 
probability, i.e. the probability to switch from the macrostate ``1'' into the
substate of the macrostate ``2'' with the conformation transition rate $\alpha_1$.
The disregarding of this very important
point would result in a serious mistake as the generated trajectories would not be stationary 
\cite{Cox,Tunaley,Goychuk04} and thus would not
correspond to the stationary conformational fluctuations which are assumed
throughout this work. The issue of noise stationarity
is crucial for our consideration.
Similar to the well-known Gillespie algorithm \cite{Gillespie}, the used here stochastic algorithm
to generate the trajectory realizations of the considered stochastic process
is exact. 

Generally, the minimum of $g\left(k_2\langle \tau_2\rangle,\theta,\zeta\right)$ 
is not related to the maximum of $C_2(\theta,\zeta)$ (and thus to the maximum of the mean 
autocorrelation time $\tau_{corr}$), even if both extrema take place in the same
parameter region $\theta\ll 1, \zeta\ll 1$, compare Fig. \ref{Fig3} and  Fig. \ref{Fig4}.
Therefore, the proper criterion is not related to the mean autocorrelation time of the considered
non-Markovian process -- this being in accordance with the discussion in the subsection \ref{diffusionmodel}.
The validity of the criterion (\ref{criterion2}) is fully confirmed in
Fig. \ref{Fig6}  for $\zeta=10^{-3}$ and three different values of $\theta$. Namely,
for $\theta=0.5$ the ET kinetics is close the regime of fast modulation,
whereas for $\theta=0.01$ it is close to the regime of static disorder. With a further
decrease of $\theta$ the kinetics starts actually to depart  from its quasi-static limit (see the dotted line for $\theta=0.001$ in Fig. \ref{Fig6}, even if the coefficient of variation $C_2$ and, therefore, 
the mean autocorrelation time $\tau_d$ attain
for this value of $\theta$ their maxima, cf. Fig. \ref{Fig3} ).

\section{Summary and conclusions}

Let us summarize the main results of this study. A simple model 
of rate processes with non-Markovian dynamical disorder has been
investigated where non-Markovian rate fluctuations were modeled by a renewal process of the
continuous time random walk type. A special attention has been paid to finding the
averaged dynamics with averaging done over the
{\it stationary} noise realizations. Such averaging corresponds to a stationary environment.
For the considered class of non-Markovian processes the main result is given 
in Eq. (\ref{result1}). In the simplest case of two-state rate fluctuations it reduces to 
the results in Eqs. (\ref{result2}) and (\ref{result3}). All these novel results are exact. 
The previously known results for the case of two-state Markovian fluctuations are 
reproduced in the simplest particular case which corresponds to the exponentially distributed
residence times in the trajectory description of the considered stochastic process.
This provides a validity test for the developed theory. 
The general theory was applied to study two
important non-Markovian models: the model incorporating a conformational diffusion within
one of the macroconformations over the energetically quasi-degenerate microstates
and the model assuming that one of the macroconformations
consists of two subconformations only (the simplest, minimal non-Markovian model). 
The former model yields a power law distribution of the residence times in the corresponding macroconformation and the latter one 
yields a biexponential distribution of the residence times.

Within the model featuring a power law distribution, a striking paradoxical effect was revealed.
Namely, the transport
process can approximately be described by the quasi-static approximation notwithstanding
that its rate  fluctuates rapidly on the transport time scale judging from
the mean residence time intervals spent by the charge transferring macromolecule  in its corresponding macroconformations. 
A proper explanation of this paradoxical effect is given
in terms of the bursting features of the underlying non-Markovian fluctuations. This explanation
was confirmed and reinforced within the latter,
simplest non-Markovian model which was solved exactly and the corresponding parameter regime
was found where the discussed paradoxical effect exists as well. The proper analytical criterion
for its occurrence was obtained for both models in similar analytical forms, 
cf. Eq. (\ref{criterion}) and Eq. (\ref{criterion2}). 
In the same spirit, the corresponding criterion can be found for any other model of the residence
time distribution $\psi_2(\tau)$ with a finite mean residence time $\langle \tau_2\rangle$. 
This is so because any such distribution (more precisely, its characteristic function, or Laplace-transform) can be written in the form of 
Eq. (\ref{distribution}) with the corresponding function $g(s,\{A\})$ found, which depends on the set 
$\{A\}$ of the model parameters such that $g(0,\{A\})=1$. If there exists such a particular parameter set 
$\{A_0\}$ that $g(k_2,\{A_0\})\ll 1$ for $k_2 \langle \tau_2\rangle \ll 1$ and $k_1 \langle \tau_1\rangle \ll 1$, then our theory predicts the occurrence of the discussed paradox of non-Markovian dynamical
disorder for this particular parameter set, cf. Eq. (\ref{result40}) and the corresponding discussion 
below Eq. (\ref{result40}).  

In practice, the paradoxical
effect revealed
in this work implies that if an ensemble measurement indicates that a charge
transfer (or another relaxation process) takes place in an apparently 
quasi-static, biexponential regime this does not mean yet
that we are dealing indeed with an ensemble of ET complexes ``frozen" randomly
in one of the
two different conformations. The following situation is possible: one-half of 
all ET complexes is indeed ``frozen" {\it temporally} in one of the 
conformations while another
half (the proportions here can be different) fluctuates very rapidly between different
conformations. 
In such a situation, the repeated
 measurements only done
with {\it single} molecules can help to uncover their actual
conformational dynamics. This would require but a sufficient time resolution of the
measurement apparatus since otherwise the most of switching effects can 
be missed
merging in an apparently long sojourn without resolving its actual fine 
structure.      

In conclusion, our study reveals that the 
non-Markovian effects can have a profound impact 
on the rate processes with dynamical disorder. Their role is expected to
be essential for the charge transfer processes in macromolecular
complexes with an intrinsic conformational dynamics. 

{\it Acknowledgments.}  This work has been supported by the Deutsche
Forschungsgemeinschaft via the Sonderforschungsbereich SFB-486,
{\em Manipulation of matter on the nanoscale}, project No. A10.

\newpage
\appendix

\section{\label{Appendix1}A model of conformational (sub)diffusion}

The conformational subspace $x$ of the state ``2'' is modeled by an one-dimensional box of the length $L$,
$-L<x<0$. The conformational
diffusion within this box is assumed to be described the fractional diffusion equation (see, e.g.,
 a review \cite{MetzlerKlafter}
and a popular account \cite{Sokolov}, as well as the further references therein). The fractional
diffusion equation can be written in the form of
standard continuity equation for the probability density $P(x,t)$,
\begin{eqnarray}\label{contin}
\frac{\partial P(x,t)}{\partial t}=-\frac{\partial J(x,t)}{\partial x}\,,
\end{eqnarray}
with the probability flux $J(x,t)$ being modified due to the
fractional time-derivative,
\begin{eqnarray}\label{flux}
J(x,t)=-K_{\alpha}  \sideset{_0}{_t}
{\mathop{\hat D}^{1-\alpha}}
\frac{\partial  P(x,t')} {\partial x} \;.
\end{eqnarray}
In (\ref{flux}), $\sideset{_0}{_t}{\mathop{\hat D}^{1-\alpha}}
(...)=\frac{1}{\Gamma(\alpha)}\frac{\partial}{\partial
t}\int_{0}^{t}\frac{(...)dt'}{(t-t')^{1-\alpha}}$ is the
integro-differential  operator of fractional derivative introduced
by Riemann and Liouville \cite{MetzlerKlafter,Sokolov} and $K_{\alpha}$ is the fractional diffusion
coefficient. For free space, $-\infty <x< \infty$, the above fractional diffusion equation
describes an anomalous subdiffusion, $\langle x^2(t)\rangle \propto t^\alpha$, with the index of 
subdiffusion $0<\alpha<1$ \cite{MetzlerKlafter,Sokolov}.
For $\alpha=1$, it reduces to the conventional diffusion equation. In the present case of the
conformation diffusion restricted in space, the diffusional time $\tau_d:=(L^2/K_\alpha)^{1/\alpha}$
becomes an important model parameter. 

Furthermore, the left boundary $x=-L$ is assumed to be reflecting,
$J(-L,t)=0$, while the right boundary is radiative, $J(0,t)= L P(0,t)/\langle \tau_2\rangle$
(here the transition into the state ``1'' occurs with the effective rate $\gamma_2=\langle \tau_2\rangle^{-1}$). Since the backward transition from the state ``1'' to the state ``2'' occurs
by crossing this right boundary in the back direction, to obtain the corresponding residence time distribution $\psi_2(\tau)$
one must solve first the fractional diffusion equation with the stated boundary conditions 
complemented by the initial condition $P(x,0)=\delta(x-x_0)$ with $x_0\to 0_{-}$. Then, the RTD
$\psi_2(\tau)$ follows as $\psi_2(\tau)=-\frac{d}{d\tau}\int_{-L}^0 P(x,\tau)dx$. This task
has been accomplished in Ref. \cite{new} using the Laplace-transform method. The result is
given in Eq. (\ref{distribution}). An analogous problem in the case of normal diffusion
was solved in Ref. \cite{PNAS02}.

The used here continuous diffusion model  has been derived in Ref. \cite{new} from a CTRW generalization of the discrete state diffusion model originally proposed by Millhauser, 
Salpeter, and Oswald \cite{Millhauser} to explain some features of the conformational gating dynamics of biological ion channels.
 It can serve, however, to model  the conformational
diffusion in other proteins as well.

\section{\label{Appendix2}Polynomial coefficients and the rates $\Gamma_i$}

In this Appendix, the exact solution of the model of
subsection \ref{exactmodel} is presented. Namely, the polynomial coefficients in Eq.
(\ref{res1biexp}) are:
\begin{eqnarray}\label{polyn1}
a_1 &=& k_2+\frac{k_1\langle \tau_2\rangle +k_2\langle \tau_1\rangle }{\langle \tau_1\rangle+
\langle \tau_2\rangle }+ \frac{1}{\langle \tau_1\rangle }+\frac{1}{\langle \tau_2\rangle }
+\frac{1}{\langle \tau_2\rangle }f_1(\theta,\zeta),\\
a_0 & = & k_2 \left( \frac{k_1\langle \tau_2\rangle +k_2\langle \tau_1\rangle }{\langle \tau_1\rangle+
\langle \tau_2\rangle }+ \frac{1}{\langle \tau_1\rangle }+\frac{1}{\langle \tau_2\rangle }\right)
\nonumber \\
& + & \frac{1}{\langle \tau_2\rangle }\frac{k_1\langle \tau_2\rangle +k_2\langle \tau_1\rangle }{\langle \tau_1\rangle+
\langle \tau_2\rangle }f_1(\theta,\zeta) +\frac{1}{\langle \tau_2\rangle }\left( \frac{1}{\langle \tau_1\rangle }+\frac{1}{\langle \tau_2\rangle }\right)f_2(\theta,\zeta),\\
b_2 &= & 2k_2 + k_1 +\frac{1}{\langle \tau_1\rangle }+\frac{1}{\langle \tau_2\rangle }
+\frac{1}{\langle \tau_2\rangle }f_1(\theta,\xi),\\
b_1 &=& k_2^2+ 2k_2 \left(k_1 + \frac{1}{\langle \tau_1\rangle } \right) +\frac{1}{\langle \tau_2\rangle }
(k_1+k_2)\left(1+ f_1(\theta,\zeta)\right) \\
&+ &\frac{1}{\langle \tau_2\rangle }\left( \frac{1}{\langle \tau_1\rangle }+\frac{1}{\langle \tau_2\rangle }\right)f_2(\theta,\zeta),\nonumber\\
b_0 &=& k_2^2 \left(k_1 + \frac{1}{\langle \tau_1\rangle } \right) +\frac{1}{\langle \tau_2\rangle }
k_1k_2\left(1+ f_1(\theta,\zeta)\right)\nonumber \\
& + &\frac{1}{\langle \tau_2\rangle }\left( \frac{k_2}{\langle \tau_1\rangle }+\frac{k_1}{\langle \tau_2\rangle }\right)f_2(\theta,\zeta).
\end{eqnarray}\label{polyn2}
In Eqs. (\ref{polyn1})-(\ref{polyn2}), two auxiliary functions,
\begin{eqnarray}\label{aux2}
f_1(\theta,\zeta)& = & (1-\theta)\zeta+\frac{\theta}{\zeta},\nonumber \\
f_2(\theta,\zeta)& = & \left((1-\theta)\sqrt{\zeta}+\frac{\theta}{\sqrt{\zeta}}\right)^2,
\end{eqnarray}
are introduced.  Furthermore, $\zeta:=\alpha_1/\alpha_2$ is the ratio of two ``on''-rates,
corresponding to the two subconformations within the macroconformation ``2'' 
 and the  mean residence time $\langle \tau_2\rangle $ is given in Eq. (\ref{MRT2}).

The rates $\Gamma_i$ in Eq. (\ref{3evolv}) are given by the roots (with the change of the sign) 
of the cubic equation which is obtained
by setting the denominator
in Eq. (\ref{res1biexp}) equal to zero. These rates read explicitly, by use of
the well-known trigonometric solution of the cubic equation\cite{Korn}:
\begin{eqnarray}\label{3rates}
\Gamma_1 &=& \frac{1}{3}b_2-2\sqrt{Q}\cos\left( \frac{\eta}{3}\right), \\
\Gamma_2 &=& \frac{1}{3}b_2-2\sqrt{Q}\cos\left( \frac{\eta+2\pi}{3}\right),\\
\Gamma_3 &=& \frac{1}{3}b_2-2\sqrt{Q}\cos\left(
\frac{\eta+4\pi}{3}\right),\label{lastrate}
\end{eqnarray}
where 
\begin{eqnarray}
\eta & = &\arccos\left ( -\frac{R}{Q^{3/2}}\right),\\
Q  &= & (b_2/3)^2 - b_1/3,\\
R  &=&  (b_2/3)^3 - b_2b_1/6+b_0/2,
\end{eqnarray}
and one assumes $Q>0$ and $|R|<Q^{3/2}$.

\clearpage



\begin{figure}
\caption{A
  sample trajectory of the considered non-Markovian process.
  When the ET complex is prepared in the excited donor state $D^*$ at $t_0=0$,
  the initial conformation $j$ (here $j=1$) was already occupied 
  before $t_0$ for some
  unknown time $\tau^*_j$. The conditioning on and the averaging
  over  $\tau^*_j$ yield generally a
  different RTD $\psi^{(1)}_j(\tau)$ 
  of the first sojourn in 
  the corresponding initial conformation. For the strictly exponential 
  $\psi_j(\tau)$ only (Markovian case), $\psi^{(1)}_j(\tau)=\psi_j(\tau)$.
  }
 \label{Fig1}
\end{figure} 

\begin{figure}
 \caption{
 The stochastically averaged decay of the donor population 
  $\langle p_D(t)\rangle_{st}$ is plotted versus the time $t$ (in arbitrary units) 
  for several different
  values of the conformational diffusion time $\tau_d$. An increase of $\tau_d$
 (from bottom to top) causes transition from the single-exponential
 ET regime of fast modulation (solid bold line, $\tau_d=0$) to a non-exponential ET regime of
  the quasi-static disorder (dashed bold line). The mean residence times in the
 conformations corresponding to the rate $k_1=1$ and the rate $k_2=0.1$ are $\langle \tau_1\rangle=0.1$
 and $\langle \tau_2\rangle=0.1$, respectively. 
 }
 \label{Fig2}
\end{figure}

\begin{figure}
 \caption{ 
 The coefficient of variation $C_2$ of the biexponential residence time
 distribution (\ref{biexp}) is plotted versus $\theta$ for different ratios $\zeta=\alpha_1/\alpha_2$
 of the rate constants $\alpha_1$ and $\alpha_2$. The dotted line corresponds to the approximation in
 Eq. (\ref{C2approx}).
 }
 \label{Fig3}
\end{figure}


\begin{figure}
 \caption{ 
 The criterion function $g(z,\theta,\zeta)$ in Eq. (\ref{gnew}) 
 is plotted versus $\theta$ for different ratios   $\zeta=\alpha_1/\alpha_2$ and for the fixed value $z=0.04$. 
 }
 \label{Fig4}
\end{figure}

\begin{figure}
 \caption{Sample trajectories of the two-state ET rate fluctuations: (a)-(c)
  and (e) corresponds to
  non-Markovian rate fluctuations with the
 single exponential distribution of the residence times in the state ``1'' with  MRT 
 $\langle \tau_1\rangle=0.2$ and with the biexponential distribution (\ref{biexp}) of 
 the residence times in the state ``2'' with  MRT 
 $\langle \tau_2\rangle=0.2$; $\zeta=0.001$ and $\theta=0.01$ (in (a) and (b)) versus $\theta=0.5$
 in (c) and (e). In (d) and (f) a sample realization of the Markovian process with the same
 MRTs is depicted on two different time scales.  }
 \label{Fig5}
\end{figure}

\begin{figure}
 \caption{ 
 The stochastically averaged decay of the donor population 
  $\langle p_D(t)\rangle_{st}$ is plotted versus the time $t$ (in arbitrary units) for the
  simplest non-Markovian model with the biexponential distribution (\ref{biexp}). 
  The solid bold line corresponds to the ET regime of fast modulation and the dashed bold
  line corresponds to the regime of quasti-static disorder. The actual ET kinetics lies
  in between these two limits. The following parameters are used in calculations: 
  $k_1=1$, $k_2=0.2$ and $\langle \tau_1\rangle=\langle \tau_2\rangle=0.2$, $\zeta=0.001$.
  For $\theta=0.01$ the ET kinetics is more close to the regime of quasi-static disorder and
  for $\theta=0.5$ the ET kinetics is more close to the fast modulation regime. 
  }
 \label{Fig6}
\end{figure} 

\clearpage

\thispagestyle{empty}
\epsfig{file=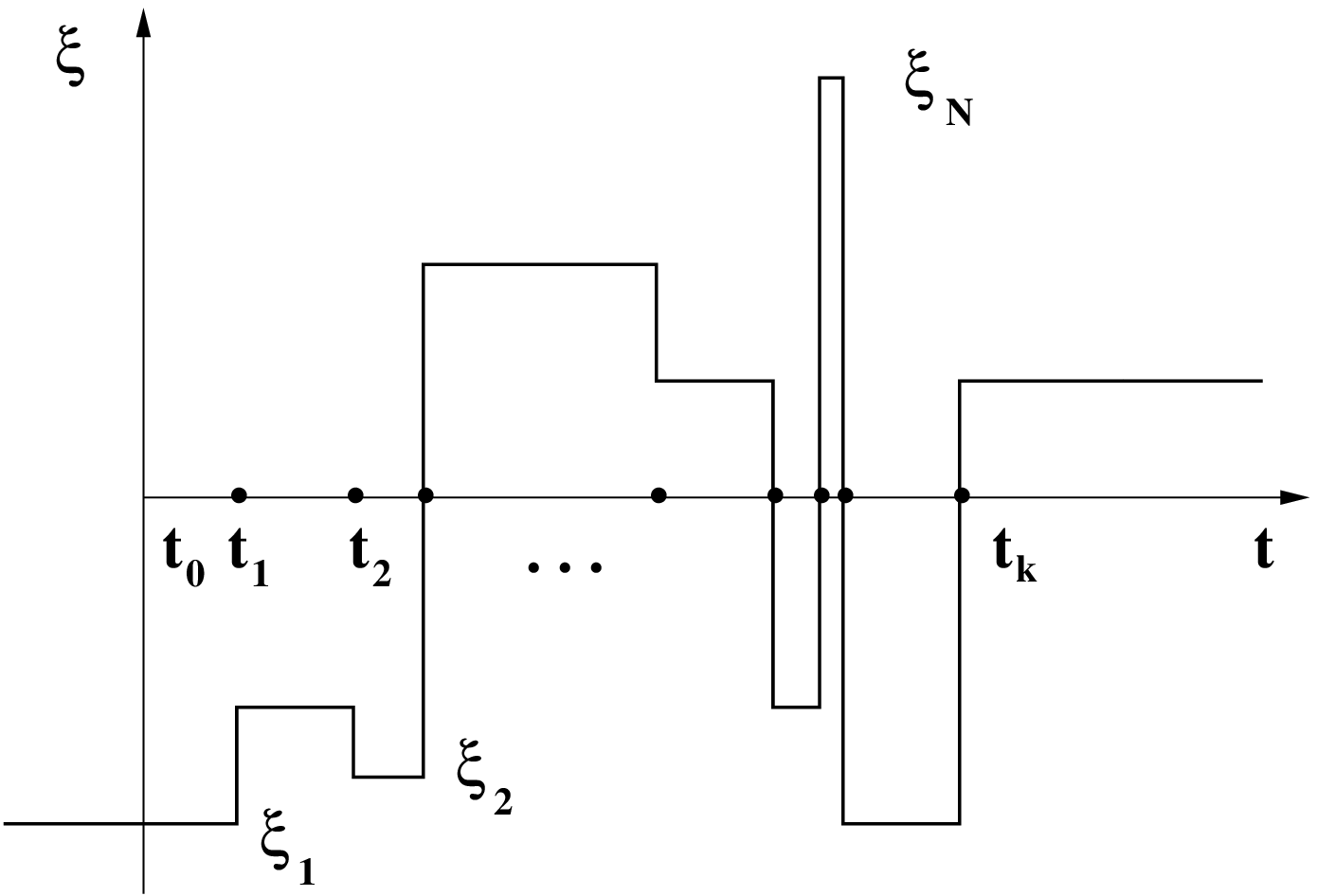,width=1\textwidth}
 \begin{center}
 FIG. 1
  \end{center}

\pagebreak
\newpage 

\thispagestyle{empty}


\epsfig{file=506517JCP2.eps,width=1\textwidth}
  \begin{center}
 FIG. 2 
 \end{center}
 
\pagebreak
\newpage

\thispagestyle{empty}

\epsfig{file=506517JCP3.eps,width=1\textwidth}
  \begin{center}
 FIG. 3 
 \end{center}
 
\pagebreak
\newpage

\thispagestyle{empty}

\epsfig{file=506517JCP4.eps,width=1\textwidth}
  \begin{center}
 FIG. 4 
 \end{center}
 

\pagebreak
\newpage

\thispagestyle{empty}

\epsfig{file=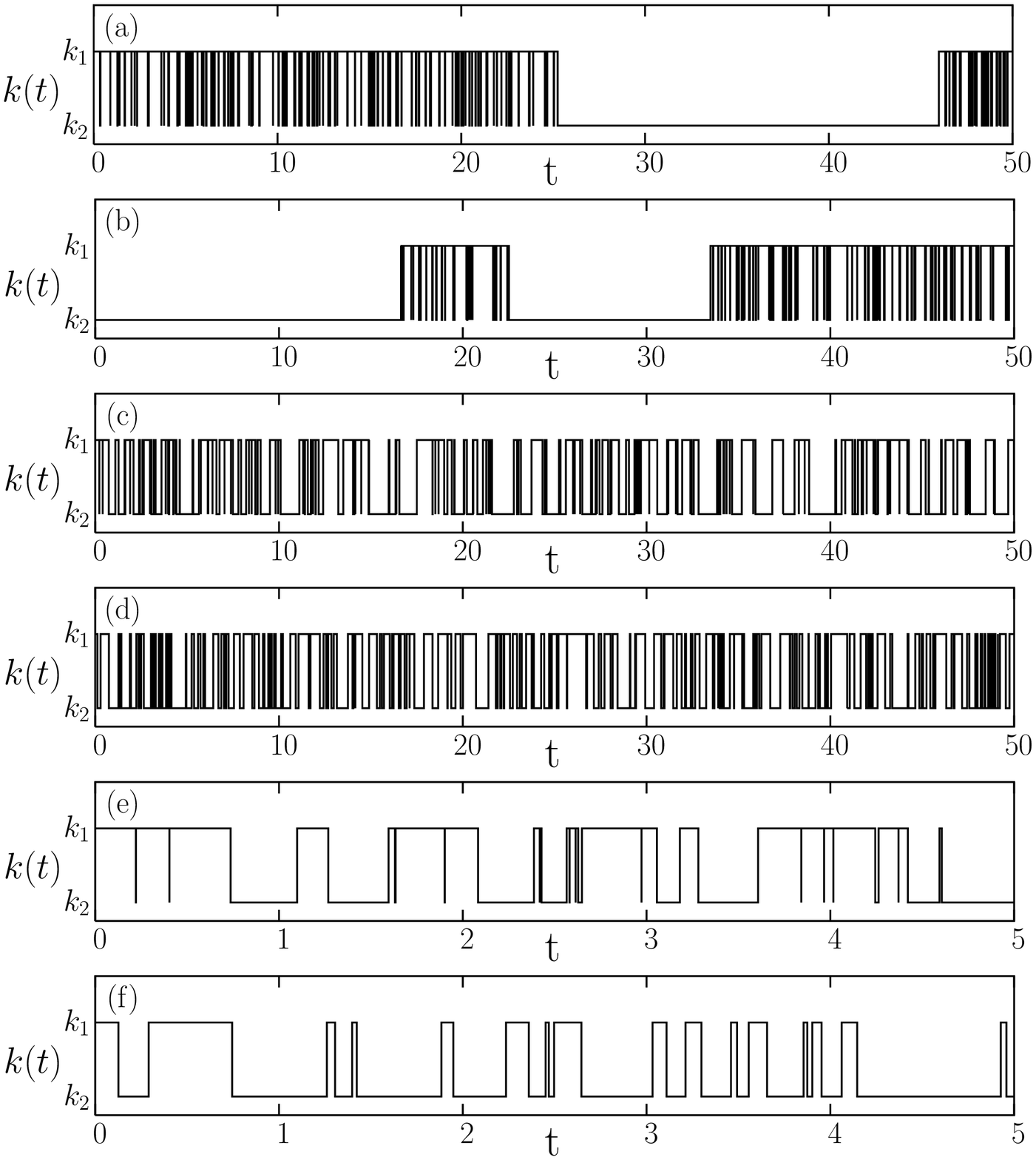,width=1\textwidth}
  \begin{center}
 FIG. 5
 \end{center}
 

\pagebreak
\newpage

\thispagestyle{empty}


\epsfig{file=506517JCP6.eps,width=1\textwidth}
  \begin{center}
 FIG. 6
 \end{center}

\end{document}